\documentclass[journal]{IEEEtran}

\ifCLASSINFOpdf
\else
   \usepackage[dvips]{graphicx}
\fi
\usepackage{url}

\usepackage{graphicx}

\usepackage{amssymb}

\usepackage{amsfonts}
\usepackage{booktabs}
\usepackage{amsmath}
\usepackage{amsthm}
\usepackage{color}
\usepackage{multirow}
\usepackage{cite}
\usepackage{makecell}
\usepackage{url}
\usepackage[colorlinks,linkcolor=red]{hyperref}

\begin{document}
\title{Local Information Assisted Attention-Free Decoder for Audio Captioning}
\author{Feiyang Xiao, Jian Guan, \IEEEmembership{Member, IEEE}, Haiyan Lan,  Qiaoxi Zhu, \IEEEmembership{Member, IEEE}, \\ 
and Wenwu Wang, \IEEEmembership{Senior Member, IEEE} 
\thanks{This work was partly supported by the Natural Science Foundation of Heilongjiang Province under Grant No. YQ2020F010, and a Newton Institutional Links Award from the British Council with Grant No. 623805725. For the purpose of open access, the authors have applied a Creative Commons Attribution (CC BY) licence to any Author Accepted Manuscript version arising. (Corresponding author: Jian Guan)}
\thanks{F. Xiao, J. Guan, and H. Lan are with the Group of Intelligent Signal Processing (GISP), College of Computer Science and Technology, Harbin Engineering University, Harbin 150001, China (emails: xiaofeiyang128@gmail.com; \{j.guan;   lanhaiyan\}@hrbeu.edu.cn).}
\thanks{Q. Zhu is with the Centre for Audio, Acoustics and Vibration, University of Technology Sydney, Ultimo, NSW, Australia (email: qiaoxi.zhu@gmail.com).}
\thanks{W. Wang is with the Centre for Vision, Speech and Signal Processing, University of Surrey, Guildford GU2 7XH, U.K.  (email: w.wang@surrey.ac.uk).}
}

\maketitle

\begin{abstract}
Automated audio captioning aims to describe audio data with captions using natural language. Existing methods often employ an encoder-decoder structure, where the attention-based decoder (e.g., Transformer decoder) is widely used and achieves state-of-the-art performance. Although this method effectively captures global information within audio data via the self-attention mechanism, it may ignore the event with short time duration, due to its limitation in capturing local information in an audio signal, leading to inaccurate prediction of captions. To address this issue, we propose a method using the pretrained audio neural networks (PANNs) as the encoder and local information assisted attention-free Transformer (LocalAFT) as the decoder. The novelty of our method is in the proposal of the LocalAFT decoder, which allows local information within an audio signal to be captured while retaining the global information. This enables the events of different duration, including short duration, to be captured for more precise caption generation. Experiments show that our method outperforms the state-of-the-art methods in Task 6 of the DCASE 2021 Challenge with the standard attention-based decoder for caption generation.
\end{abstract}

\begin{IEEEkeywords}
Automated audio captioning, local information, attention-free Transformer 
\end{IEEEkeywords}

\IEEEpeerreviewmaketitle

\section{Introduction}
\label{sec:intro}

\IEEEPARstart{A}{utomated} audio captioning (AAC) is an intermodal translation task that converts input audio into a text description, i.e., caption, using natural language \cite{drossos2020clotho, mei2022automated}. It benefits various applications, such as intelligent and content oriented machine-to-machine interaction and automatic content description \cite{xu2021_t6, mei2022diverse, liu2022leveraging, liu2021cl4ac}. 
Existing AAC methods usually employ an encoder-decoder structure, where the encoder is used to obtain an audio feature of the audio input, while the decoder is used to generate texts from the audio feature \cite{drossos2017automated, mei2021audio, koizumi2020transformer, mei2022automated, xu2022comprehensive}. The pretrained audio neural networks (PANNs) \cite{kong2020panns} module has been widely used in the encoder to extract audio features from the audio signal, while the attention-based Transformer has been used in the decoder to model the global information within audio features with a self-attention mechanism \cite{mei2021audio}. With the PANNs encoder and the attention-based decoder, several methods \cite{xinhao2021_t6, narisetty2021_t6, chen2021_t6, xu2021_t6} have been developed, achieving state-of-the-art performance in Task 6 of the DCASE 2021 Challenge.

The standard self-attention mechanism \cite{mei2021audio}, however, does not explicitly consider the local information within audio features and may ignore the event with short duration. Moreover, with the softmax function used in the self-attention strategy, the frames that contain predicted events get high weights, while those corresponding to rare events may get low weights. As a result, the prediction of captions for describing such events may be inaccurate. Table \ref{tab:caption_eg} shows examples of generated captions using such methods, assuming the ground truth captions are correct. 
For instance, P-Transformer \cite{xinhao2021_t6} may miss local information about some events, e.g., ``\textit{people are talking}" in example 1, or generate inaccurate descriptions, e.g., ``\textit{A dog barks}" is wrongly interpreted as ``\textit{birds are chirping}" in example 2.

\begin{table}[t]
  \centering
  \caption{Examples show how the loss of local information and imprecise caption prediction degrade the quality of captions.}
    \resizebox{\columnwidth}{!}{
    \begin{tabular}{p{0.22\columnwidth}|c|p{0.6\columnwidth}}
    \hline 
    \makecell[c]{Example} & Method & \makecell[c]{Caption} \\
    \hline
    \multirow{5}{*}{\makecell[l]{1. Loss of local \\information}} & \multirow{2}{*}{Ground Truth} & Wind chimes are playing while \textit{people are talking} in the background \\
    \cline{2-3} 
    & \multirow{3}{*}{P-Transformer \cite{xinhao2021_t6}} & Wind chimes chime wind chimes chime wind chimes clang wind chimes tinkle wind chimes and wind chimes tinkle wind chimes chime \\
    \cline{2-3}
    \hline 
    \multirow{4}{*}{\makecell[l]{{2. Imprecise} \\{caption prediction}}} & \multirow{2}{*}{Ground Truth} & \textit{A dog barks} a few times and two men talk to each other \\
    \cline{2-3} 
    & \multirow{2}{*}{P-Transformer \cite{xinhao2021_t6}} & People are talking while \textit{birds are chirping} in the background \\
    \cline{2-3}
    \hline 
    \end{tabular}
    }
  \vspace{-5mm}
  \label{tab:caption_eg}
\end{table}%

To address the above issues, we propose a new decoder by leveraging the attention-free Transformer (AFT) method and its variant AFT-local \cite{zhai2021attention}. AFT-local was initially proposed for modeling data within a local region in image processing and character-level language modeling \cite{zhai2021attention}. Although AFT-local has the ability to capture locality while maintaining global connectivity within features, it does not have a decoder structure to model the sequential information for text description, which, nevertheless, is required in the audio captioning task. Other studies, e.g., \cite{liu2021swin} for image classification and \cite{han2022local} for speaker verification have also considered local self-attention, but they cannot be used directly to model the sequential information for generating text description.

In this letter, we present a LocalAFT decoding method based on \cite{zhai2021attention} for the AAC task. More specifically, our LocalAFT method has two key modules, namely, the future interference masking (FIM) module and the local information assisted captioning (LAC) module. The FIM module is designed to model the sequential word features, which can avoid interference from future words and use element-wise multiplication to capture the global information of the input word features. In the LAC module, we introduce a window function and a learnable weight matrix, where the window function is used to constrain the weights in the learnable weight matrix. With the element-wise multiplication and the constrained weight matrix, our LAC module can capture the local information while maintaining the global information of audio features, to predict latent features for captions, thereby achieving the inter-modal integration between text and audio information. Thus, the proposed decoder can capture the short-time event, along with events of longer duration, and generate more precise captions.

In our work, the PANNs algorithm based encoder is employed to extract high-quality audio features as the input of the LocalAFT decoder. Therefore, the overall captioning method including both encoder and decoder is denoted as P-LocalAFT. The proposed method is evaluated and compared with the state-of-the-art methods that use PANNs as the encoder and the standard attention mechanism as the decoder \cite{xinhao2021_t6, narisetty2021_t6, chen2021_t6, xu2021_t6}. Experiments are conducted on the AudioCaps dataset \cite{kim2019audiocaps} for pretraining the model and the Clotho-v2 dataset \cite{drossos2020clotho} for fine-tuning the model. Results show that the proposed method achieves better performance than the state-of-the-art methods.

\section{Proposed Method}
\label{sec:method}
\begin{figure}[!htbp]
  \vspace{-5mm}
  \centering
  \centerline{\includegraphics[width=0.82\columnwidth]{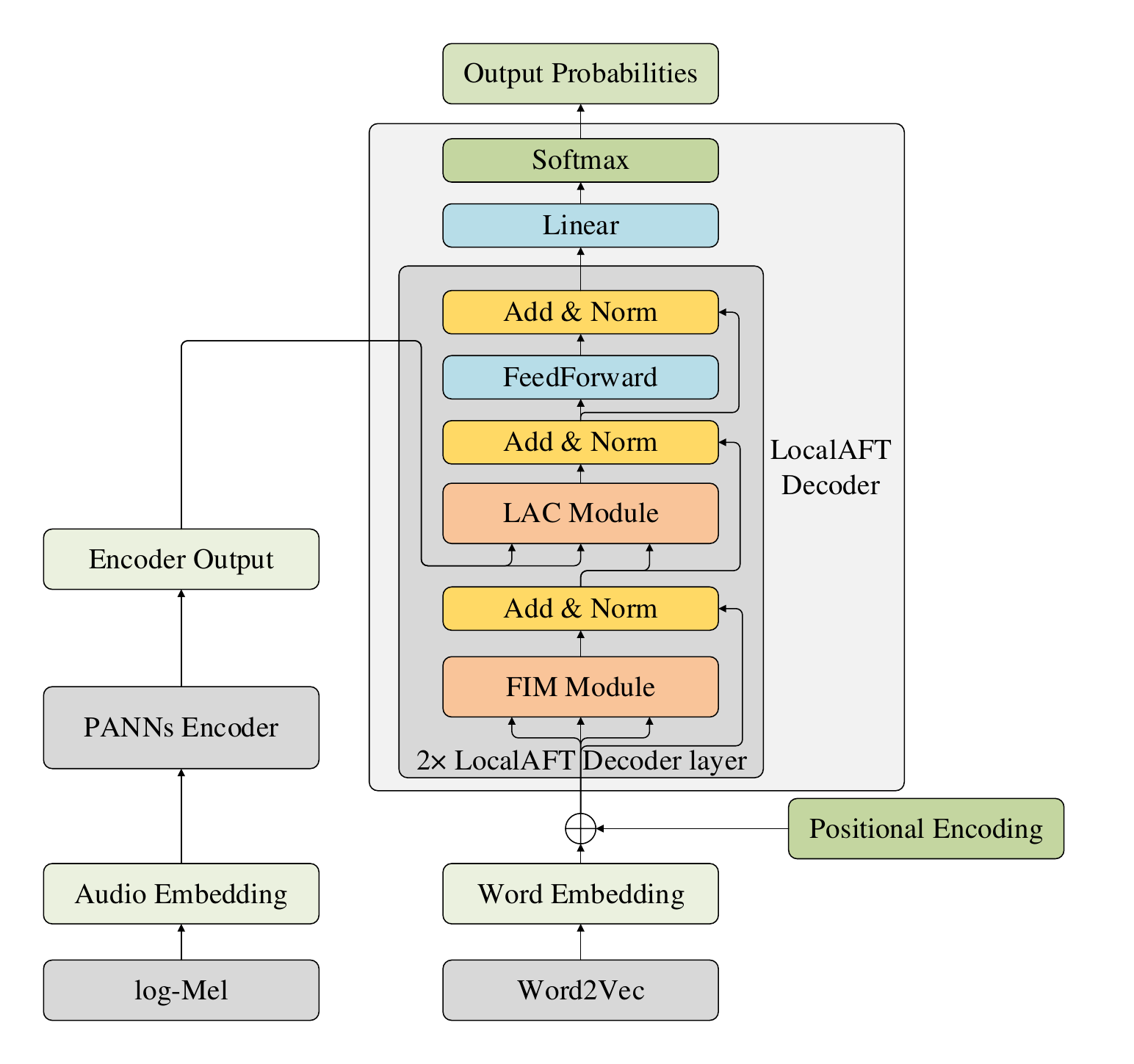}}
  \caption{The framework of the proposed automated audio captioning method.}
  \label{fig:1}
  \vspace{-5mm}
\end{figure}

The proposed P-LocalAFT uses a PANNs encoder to extract audio features and a LocalAFT decoder to incorporate the local information about audio events to guide caption prediction. Figure \ref{fig:1} shows the framework of the P-LocalAFT method. 

\subsection{PANNs Encoder}
\label{subsec:encoder}
The encoder of the proposed method is used to extract audio features, which are the input of the decoder for caption prediction. The input to the PANNs encoder is the log-Mel spectrogram $\mathbf{X} \in \mathbb{R}^{T \times F}$, where $T$ denotes the number of time frames and $F$ denotes the dimension of Mel band. 

To extract the audio feature, we choose the 10-layer CNN module from PANNs \cite{kong2020panns} which is pretrained on AudioSet \cite{gemmeke2017audio}. The 10-layer CNN module has four convolutional blocks and two linear layers. Each convolutional block has two $3 \times 3$ convolutional layers with ReLU activation function and batch normalization. The number of channels in the convolutional blocks is 64, 128, 256, and 512, respectively. There are $2 \times 2$ average pooling layers between convolutional blocks. After the convolutional blocks, global pooling is employed on the Mel band dimension. Then, two linear layers are used to output the audio feature $\mathbf{H} \in \mathbb{R}^{L \times D}$, as the input to the LocalAFT decoder, where $L$ and $D$ represent the number of time frames and the dimension of the spectral feature at each frame, respectively. Here $D$ is set as $128$. 
\vspace{-0.5mm}
\subsection{LocalAFT Decoder}
\label{subsec:decoder}
Inspired by \cite{zhai2021attention}, in our LocalAFT decoder, local information about audio events is incorporated to guide caption prediction. The LocalAFT decoder has two inputs. One is the output of the PANNs encoder $\mathbf{H} \!\in\! \mathbb{R}^{L \times D}$, the other is the word embedding $\mathbf{W} =\left[ \mathbf{w}_0, \cdots, \mathbf{w}_{N-1} \right]^\top \in \mathbb{R}^{N \times D}$ from a pretrained word2vec model, where $N$ denotes the length of the target caption and $\top$ denotes matrix transpose. Note that, $\mathbf{w}_0$ is the token $<\!\text{sos}\!>$ representing the start of a sequence, and it is not from a caption. To predict the $n$-th word $\mathbf{w}_n$ of the caption, the posterior probability is calculated as
\begin{equation}
\label{eq:decoding}
    p \left( \mathbf{w}_n | \mathbf{H}, \mathbf{W}_{pre} \right) = Decoder \left(\mathbf{H}, \mathbf{W}_{pre} \right),
\end{equation}
where $\mathbf{W}_{pre} = \left[ \mathbf{w}_0, \cdots, \mathbf{w}_{n-1} \right]^\top$ are the previously predicted words. The LocalAFT decoder has two key modules, future interference masking (FIM) and local information assisted captioning (LAC), as shown in Figure \ref{fig:1}. FIM is developed to deal with sequential word embedding and avoid the interference from the future words after the $(n-1)$-th word, i.e., $[\mathbf{w}_n, \cdots, \mathbf{w}_{N-1}]^\top$. LAC is developed to capture the local information while maintaining the global information about audio events.
\subsubsection{\textbf{Future Interference Masking}}
\label{subsubsec:maskedAFT}
This module processes the input word feature $\mathbf{Y}$ (i.e., the word embedding with positional encoding), where $\mathbf{Y}=[\mathbf{y}_1,  \cdots, \mathbf{y}_n, \cdots,  \mathbf{y}_N]^\top \!\in\! \mathbb{R}^{N \times D}$. To predict the $n$-th word, the input feature $\mathbf{Y}$ should only contain the features of the previously generated words (i.e., $[\mathbf{y}_1, \cdots, \mathbf{y}_n]^\top$), and the feature after  $\mathbf{y}_n$ should be ignored. To this end, we design a mask matrix $\mathbf{M} \in \mathbb{R}^{N \times N}$ of learnable weights in the FIM module, defined as
\begin{equation}
\label{eq:masked_tril}
    \mathbf{M} \!=\! \begin{bmatrix}
            m_{1, 1}   & -\infty    & \cdots & -\infty  \\    
            m_{2, 1}   & m_{2, 2}   & \ddots & \vdots \\   
            \vdots     & \vdots     & \ddots & -\infty \\     
            m_{N, 1}   & m_{N, 2}   & \cdots & m_{N, N}   
        \end{bmatrix},
\end{equation}
where ``$-\infty$" is the negative infinity and $\exp(-\infty)\!=\!0$. As a result, $\exp(\mathbf{M})$ becomes a lower triangular matrix. This helps reduce the interference from subsequent words, i.e., words after the current words.

Following the above masking, we define the future interference masking function $FIM(\cdot)$ to remove the future interference in $\mathbf{Y}$, as follows
\begin{equation}
\label{eq:aft_masked}
\begin{split}
 \mathbf{y}_{n}^{FIM} & \!=\! FIM\left ( \mathbf{Y} \right ) \\
 & \!=\! \sigma(\mathbf{Q}_n) \!\odot\! \frac{\sum_{i=1}^{I} \exp(m_{n,i}) \odot \exp(\mathbf{K}_{i}) \odot \mathbf{V}_{i}} {\sum_{i=1}^{I} \exp(m_{n,i}) \odot \exp(\mathbf{K}_{i})},\\
\end{split}
\end{equation}
where $\mathbf{y}_{n}^{FIM}$ denotes the output feature of the FIM module corresponding to the $n$-th word $\mathbf{w}_n$, $m_{n,i}$ is an element of matrix $\mathbf{M}$, $\sigma(\cdot)$ denotes the sigmoid function, $\odot$ denotes element-wise multiplication, and $I$ is the length of the previously generated caption. Here, $\mathbf{Q}$, $\mathbf{K}$ and $\mathbf{V}$ are the linear mapping results of $\mathbf{Y}$, which are calculated following \cite{zhai2021attention}. $\mathbf{Q}_n$ is the $n$-th row vector of $\mathbf{Q}$. $\mathbf{K}_{i}$ and $\mathbf{V}_{i}$ are the $i$-th row vectors of $\mathbf{K}$ and $\mathbf{V}$, respectively.

According to Eq. \eqref{eq:aft_masked}, the FIM module is free from the words information after the previously generated caption, since $\exp(m_{n,i})\!=\!0$ is used to mask future interference when $i>n$. The mask matrix $\mathbf{M}$ can act on all captions of the dataset and adjust the weight on its lower triangular part. Instead of implementing self-attention through matrix multiplication operation, the FIM applies element-wise multiplication operation, following \cite{zhai2021attention}. Therefore, the FIM module involves global information similar to self-attention for each caption feature, by applying $\mathbf{M}$ over the whole dataset. 

Then, we can obtain the output feature $\mathbf{\hat{y}}_{n}^{FIM}$ of the $n$-th word, as follows
\begin{equation}
\label{eq:add&norm_MFA}
    \mathbf{\hat{y}}_{n}^{FIM} = LN \left( \mathbf{y}_{n}^{FIM} + \mathbf{y}_{n} \right),
\end{equation}
where $\mathbf{y}_{n}$ is the $n$-th vector of the input feature $\mathbf{Y}$, corresponding to the $(n-1)$-th word $\mathbf{w}_{n-1}$, and $LN(\cdot)$ denotes the layer normalization function.

\subsubsection{\textbf{Local Information Assisted Captioning}}
\label{subsubsec:localAFT}
The local information assisted captioning (LAC) module uses both $\mathbf{\hat{y}}_n^{FIM}$ and audio feature $\mathbf{H}$ as inputs, aiming to learn the semantic information from the audio feature and generate latent caption feature. To learn the global information from the whole dataset, a weight matrix $\mathbf{Z} \in \mathbb{R}^{N \times L}$ is defined and learned in LAC. Meanwhile, similar as \cite{zhai2021attention}, a window function $f_{\mathcal{L}}(\cdot)$ is used to constrain the weight matrix $\mathbf{Z}$, as follows 
\begin{equation}
\label{eq:local}
\begin{split}
    f_{\mathcal{L}} \left( z_{n, l} \right) = \left\{\begin{matrix}
                                               z_{n, l}, & if \ (l-n) < s, \\
                                               0,        & \text{otherwise}, 
                                           \end{matrix}\right.  
\end{split}
\end{equation}
where $z_{n, l} \in \mathbf{Z}$ with $1 \le n \le N$ and $1 \le l \le L$. In Eq. \eqref{eq:local}, $s$ is the size of the local region, which helps $\mathbf{Z}$ attend local information in audio feature $\mathbf{H}$. Different from \cite{zhai2021attention}, the local area defined in Eq. \eqref{eq:local} is not lower bounded with respect to $s$ and $n$. Specifically, $\mathbf{Z}$ correlates the caption with the audio feature, such that the local information corresponding to the $n$-th word information can be adaptively learned from the $1$-st to the $(s+n)$-th audio feature vectors.

The output feature $\mathbf{y}_{n}^{LAC}$ corresponding to the $n$-th word $\mathbf{w}_n$ is calculated as
\begin{equation}
\label{eq:LAC}
    \begin{split}
        \mathbf{y}_{n}^{LAC} & \! = \!LAC\left ( \mathbf{\hat{y}}_n^{FIM}, \mathbf{H} \right ) \\
        & \! = \!\sigma(\mathbf{\bar{y}}_n) \!\odot\! \frac{\sum_{l=1}^{L} \exp(f_{\mathcal{L}}(z_{n,l})) \!\odot\! \exp(\mathbf{H}_{l}^K) \!\odot\! \mathbf{H}_{l}^V} {\sum_{l=1}^{L} \exp(f_{\mathcal{L}}(z_{n,l})) \!\odot\! \exp(\mathbf{H}_{l}^K)},\\
    \end{split}
\end{equation}
where $\mathbf{\bar{y}}_{n}$ is the linear mapping result of $\mathbf{\hat{y}}_{n}^{FIM}$, $\mathbf{H}^K$ and $\mathbf{H}^V$ are the linear mapping results of $\mathbf{H}$. Here, $\mathbf{H}_{l}^K$ and $\mathbf{H}_{l}^V$ are the $l$-th row vector of $\mathbf{H}^K$ and $\mathbf{H}^V$, respectively.

Note that, when $z_{n,l}$ is not within the window for the local region, $\exp(f_{\mathcal{L}}(z_{n,l}))$ has the value as 1 (i.e., $\exp(0)=1$). Therefore, $\mathbf{Z}$ can learn local information (i.e., $f_{\mathcal{L}}(z_{n,l})=z_{n,l}$), while allowing the LAC module to retain the global information by the Transformer (i.e., $f_{\mathcal{L}}(z_{n,l})=0$). In summary, the LAC module can model both the global and local information for caption prediction. Finally, the output feature of LAC is calculated as follows
\begin{equation}
\label{eq:add&norm_LAC}
    \mathbf{\hat{y}}_{n}^{LAC} = LN \left( \mathbf{y}_{n}^{LAC} + \mathbf{\hat{y}}_{n}^{FIM} \right).
\end{equation}

To get a more effective word embedding, we pretrain a word2vec language model \cite{mikolov2013efficient} by combining the captions of both Clotho-v2 \cite{drossos2020clotho} and an external AAC dataset (i.e., AudioCaps \cite{kim2019audiocaps}). We also use the AudioCaps dataset to pretrain the whole P-LocalAFT method, and then fine-tune it on Clotho-v2 dataset, with the same training strategy as \cite{xinhao2021_t6}.

\section{Experiments}
\label{sec:experiment}
\subsection{Dataset}
\label{subsec:dataset}
We use two datasets in our experiments, i.e., Clotho-v2 \cite{drossos2020clotho} and AudioCaps \cite{kim2019audiocaps}. Clotho-v2 was released for Task 6  of the DCASE 2021 Challenge, including 3839, 1045 and 1045 audio clips for the development, validation and evaluation splits, respectively. Each audio clip has five captions. The audio clips have a varying duration ranging from 15 to 30 seconds, with captions of 8 to 20 words. In our experiments, the development and validation splits are combined together to form a new training set of 4884 audio clips. The evaluation split is used for the performance evaluation. 

AudioCaps \cite{kim2019audiocaps} is another AAC dataset whose audio signals are from AudioSet \cite{gemmeke2017audio}, including 49838, 495 and 975 audio clips for the development, validation and evaluation splits, respectively, in the initial version. However, because of the rules of YouTube, the dataset available in this experiment includes 44366, 458 and 905 audio clips for the development, validation and evaluation, respectively. Each audio clip has the duration of 10 seconds. We combine the development and the validation splits to pretrain our P-LocalAFT method. 

\begin{table*}[t]
  \centering
  \caption{Performance comparison on the evaluation split of the Clotho-v2 dataset.}
  \resizebox{0.98\textwidth}{!}
  {
    \begin{tabular}{cccccccccc}
    \toprule
    Method & \multicolumn{1}{l}{BLEU$_1$} & \multicolumn{1}{l}{BLEU$_2$} & \multicolumn{1}{l}{BLEU$_3$} & \multicolumn{1}{l}{BLEU$_4$} & \multicolumn{1}{l}{ROUGE$_l$} & \multicolumn{1}{l}{METEOR} & \multicolumn{1}{l}{CIDE$_r$} & \multicolumn{1}{l}{SPICE} & \multicolumn{1}{l}{SPIDE$_r$} \\
    \midrule
    P-Transformer \cite{xinhao2021_t6} & 0.565 & 0.372 & 0.253 & 0.171 & 0.377 & 0.172 & 0.413 & \textbf{0.123} & 0.268 \\
    P-Conformer \cite{narisetty2021_t6} & 0.546 & 0.356 & 0.243 & 0.165 & 0.369 & 0.163 & 0.381 & 0.110  & 0.246 \\
    P-Meshed Memory \cite{chen2021_t6} & 0.563 & 0.367 & 0.244 & 0.158 & 0.371 & 0.170 & 0.406 & 0.119  & 0.262 \\
    P-Temporal Attention \cite{xu2021_t6} & 0.576 & 0.377 & 0.252 & 0.164 & 0.382 & \textbf{0.178} & 0.421 & 0.122 & 0.271 \\
    \midrule
    \textbf{P-GlobalAFT} & 0.563 & 0.374 & 0.255 & 0.171 & 0.374 & 0.171 & 0.414 & 0.120  & 0.267 \\
    \textbf{P-LocalAFT} & \textbf{0.578} & \textbf{0.387} & \textbf{0.267} & \textbf{0.179} & \textbf{0.390}  & 0.177 & \textbf{0.434} & 0.122 & \textbf{0.278} \\
    \bottomrule
    \end{tabular}
  }
  \label{tab:1}
  \vspace{-0.5cm}
\end{table*}
\subsection{Experimental Setup}
\label{subsec:setup}
The local region window size $s$ in Eq. \eqref{eq:local} was set as 80 based on empirical tests. In practice, this hyper-parameter needs to be chosen and tuned in terms of the data to be processed. Positional encoding and padding mask were used in the LocalAFT decoder. For the whole P-LocalAFT method, SpecAugment and mix-up strategies were introduced to improve generalization following \cite{xinhao2021_t6}. The cross-entropy loss with label smooth \cite{szegedy2016rethinking} was used with Adam optimizer \cite{kingma2014adam} to optimize the network. The batch size was set as 16, and the learning rate was set as 0.0001. The decoder used a teacher forcing strategy in training and beam search strategy with the beam size of 5 in evaluation.

Following DCASE 2021 Challenge, all the methods are evaluated by machine translation metrics (i.e., BLEU$_n$, ROUGE$_l$ and METEOR) and captioning metrics (CIDE$_r$, SPICE and SPIDE$_r$). BLEU$_n$ \cite{papineni2002bleu} measures a modified n-gram precision. ROUGE$_l$ \cite{lin2004rouge} is a score based on the longest common sub-sequence. METEOR \cite{lavie2007meteor} is a harmonic mean of weighted unigram precision and recall. CIDE$_r$ \cite{vedantam2015cider} is a weighted cosine similarity of n-grams, which reflects the grammar and fluency properties of captions ignored in METEOR and SPICE. SPICE \cite{anderson2016spice} is the F-score of semantic propositions extracted from caption and reference. SPIDE$_r$ \cite{liu2017improved} is the mean score between CIDE$_r$ and SPICE.
\subsection{Performance Comparison}
\label{subsec:results}
The proposed P-LocalAFT is compared with state-of-the-art methods \cite{xinhao2021_t6, narisetty2021_t6, chen2021_t6, xu2021_t6}, which all use PANNs encoder and standard attention decoders, i.e., P-Transformer \cite{xinhao2021_t6}, P-Conformer \cite{narisetty2021_t6}, P-Meshed Memory \cite{chen2021_t6}, and P-Temporal Attention \cite{xu2021_t6}. Noting that, our proposed method does not consider the influence of reinforcement learning, so the reinforcement learning processes in both P-Transformer and P-Temporal Attention are not involved. Moreover, the P-Transformer method has the same training strategy as the proposed P-LocalAFT. The results of P-Conformer, P-Meshed Memory, and P-Temporal Attention are those from the results of DCASE 2021 Challenge.

The results are shown in Table \ref{tab:1}. Our proposed P-LocalAFT method outperforms other state-of-the-art methods (i.e., P-Transformer, P-Conformer, P-Meshed Memory and P-Temporal Attention), in terms of BLEU$_n$, ROUGE$_l$, CIDE$_r$ and SPIDE$_r$. More specifically, apart from the SPICE metric, our P-LocalAFT method performs better than P-Transformer in all the other metrics, showing that our LocalAFT decoder outperforms the standard Transformer decoder, especially in terms of grammar and fluency. The proposed method has slightly lower performance in SPICE and METEOR as compared to the best baselines. This could be because the window size is fixed and not optimized for capturing the semantic information of the audio events or scenes with varying duration.

\subsection{Ablation Study}
\label{subsec:ablation}
To demonstrate the effectiveness of using local information for audio captioning, a P-LocalAFT variant without using the local region (i.e.,  P-GlobalAFT) is evaluated in our ablation study. Specifically, P-GlobalAFT does not use the local region window function (i.e., Eq. \eqref{eq:local}) to constrain the learnable weight matrix $\mathbf{Z}$ in the LAC module. The result is also given in Table \ref{tab:1}. It can be observed from Table \ref{tab:1} that, without incorporating the local information, P-LocalAFT degenerates to P-GlobalAFT, resulting in similar captioning performance to the standard Transformer-based method, i.e., P-Transformer. This result indicates that the proposed P-LocalAFT improves performance in caption prediction, due to the effective use of the local information from the audio events and caption contents, along with the use of the global information.
\begin{table}[t]
  \centering
    \caption{Illustration for audio caption using the methods with and without local information.}
    \resizebox{\columnwidth}{!}{
    \begin{tabular}{p{0.22\columnwidth}|c|p{0.6\columnwidth}}
    \hline 
    \makecell[c]{Example} & Method & \makecell[c]{Caption} \\
    \hline
    \multirow{9}{*}{\makecell[l]{1. Loss of local \\information}} & \multirow{2}{*}{Ground Truth} & Wind chimes are playing while \textit{people are talking} in the background \\
    \cline{2-3} 
    & \multirow{3}{*}{P-Transformer \cite{xinhao2021_t6}} & Wind chimes chime wind chimes chime wind chimes clang wind chimes tinkle wind chimes and wind chimes tinkle wind chimes chime \\
    \cline{2-3}
    & \multirow{2}{*}{P-GlobalAFT} & Wind chimes clang together in the wind as the wind blows \\
    \cline{2-3}
    & \multirow{2}{*}{\textbf{P-LocalAFT}} & Wind chimes clang together as \textit{people talk} in the background \\
    \hline 
    \multirow{8}{*}{\makecell[l]{{2. Imprecise} \\{caption prediction}}} & \multirow{2}{*}{Ground Truth} & \textit{A dog barks} a few times and \textit{two men talk} to each other \\
    \cline{2-3} 
    & \multirow{2}{*}{P-Transformer \cite{xinhao2021_t6}} & \textit{People are talking} while \textit{birds are chirping} in the background \\
    \cline{2-3}
    & \multirow{2}{*}{P-GlobalAFT} & \textit{A dog barks} while \textit{birds chirp} in the background \\
    \cline{2-3}
    & \multirow{2}{*}{\textbf{P-LocalAFT}} & \textit{A dog barks} while \textit{people talk} in the background \\
    \hline 
    \end{tabular}
    }
  \vspace{-5mm}
  \label{tab:caption_comparison}%
\end{table}%

For better understanding the effect of local information for audio captioning, we present two examples in Table \ref{tab:caption_comparison}. In example 1, the audio event ``\textit{people talk}'' is described by our P-LocalAFT method but lost in the caption generated by P-Transformer and P-GlobalAFT. This demonstrates that our LocalAFT decoder can capture the audio event of short duration, overcoming the loss of local information in the standard Transformer decoder. Meanwhile, in example 2, the audio event ``\textit{a dog barks}" is wrongly interpreted as ``\textit{birds are chirping}" by the P-Transformer method and the audio event ``\textit{two men talk}" is wrongly interpreted as ``\textit{birds chirp}" by the P-GlobalAFT method, which shows that the lack of the effective use of the local information may also degrade the quality of the prediction of the acoustic events. In contrast, the proposed P-LocalAFT method precisely describes the events ``\textit{a dog barks}" and ``\textit{people talk}" in the generated caption, verifying that the proposed LocalAFT decoder can achieve more precise caption prediction. The source codes and more examples with audio clips are available \footnote{\url{https://github.com/LittleFlyingSheep/P-LocalAFT/tree/main/examples}}.
\section{Conclusion}
\label{sec:conclusion}
We have presented a novel automated audio captioning method that employs a PANNs encoder to extract audio features and exploits a LocalAFT decoder to incorporate local and global information from audio data. Experimental results show that the proposed method improves captioning performance as compared to state-of-the-art methods by incorporating local information, and overcomes the limitation of these attention-based methods. In a future work, we may explore the choice of adaptive window size to capture local information from audio clips with acoustic events and scenes of different duration.
%
%


\end{document}